\documentclass[11pt,a4paper]{article}
\usepackage{authblk}

\usepackage{graphicx,amsmath,bm}

\begin{document}

\title{Coherent perfect absorption mediated enhancement and optical bistability in phase conjugation.}
\author[1]{K. Nireekshan Reddy}
\author[2]{Achanta Venu Gopal}
\author[1,*]{S. Dutta Gupta}
\affil[1]{School of Physics, University of Hyderabad, Hyderabad-500046, India}
\affil[2]{DCMPMS, Tata Institute of Fundamental Research, Mumbai-400005, India}
\affil[*]{email: \textit{sdghyderabad@gmail.com}}
\renewcommand\Authands{ and }


\maketitle



\begin{abstract}
We study  phase conjugation in a nonlinear composite slab when the counter propagating pump waves are completely absorbed by means of coherent perfect absorption. Under the undepleted pump approximation the coupling constant and the phase conjugated reflectivity are shown to undergo a substantial increase and multivalued response. The effect can be used for efficient switching of the phase conjugated reflectivity in photonic circuits.  
\end{abstract}






\section{Introduction}The ability of light waves to interfere destructively or constructively has found several applications in recent years. Some of the most notable applications are critical coupling and coherent perfect absorption \cite{Yariv-cc1,Bulovic,sdg-cc,deb1,deb2,deb3,Wan-sc,opex-12,bacbendl,knr-sdg-rev}. The applications are not limited to just manipulation of intensities, similar ideas are extended to intensity-intensity correlations in quantum optics \cite{HUM}. It has been shown recently that such destructive interference in higher order correlations can lead to a perfect Hong-Ou-Mandel dip with 100\% visibility \cite{sdg-gsa-14}, with values as high as 95\% reported in recent experiments \cite{Atwater}. In the context of standard critical coupling (CC) and coherent perfect absorption (CPA) the remarkable possibility of controlling absorption has been demonstrated convincingly in both theory and experiments. The recent efforts are aimed at making use of the CC and CPA phenomenon for other applications. One of the applications that comes to the fore are different nonlinear effects \cite{knr1,knr2,knr3}. In nonlinear optics, one mostly focuses on how to enhance the nonlinear interactions and thereby the efficiency of the outgoing waves. Indeed, only part of the interacting pump waves in harmonic generation, or four-wave mixing and phase conjugation is utilized for effective nonlinear conversion. CPA and CC open new possibilities that these waves can be completely absorbed by nonlinear medium. Such initial trends have been probed by several others in the context of harmonic generation \cite{np-lon-cpa,lon-choas}. Our present investigation focuses on combining the concepts of CPA and phase conjugation. In an earlier paper, we have reported CPA in a nonlinear medium to show bistable and multistable response \cite{knr2}. We extend those results to the pump waves to evaluate the coupling coefficient for the phase conjugation process. In the undepleted pump approximation, coupling parameters determine the evolution of the signal and the phase conjugated waves \cite{yariv-pc,boyd}. We show that there is a drastic change in the phase conjugated reflectivity at each of the nonlinear CPA resonances. The bistability in the pump reflects into bistability of the coupling parameters and the phase conjugated reflectivity. The control on the coupling parameter by means of pump intensity renders the oscillation condition to be power dependent. One can thus have oscillation condition for phase conjugation fulfilled at much lower system size.
\section{Phase conjugation with CPA of pump waves}
\begin{figure}[h]
\center{\includegraphics[width=5.25cm]{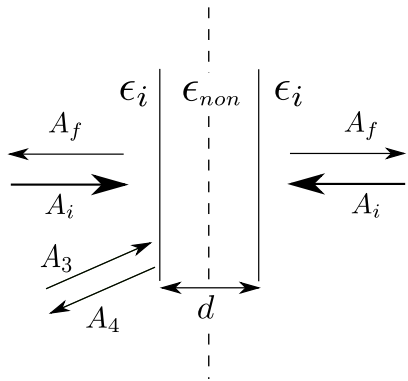}}
\caption{Schematic view of nonlinear slab with thickness $d$ under CPA geometry (with identical pump powers $|A_i|^2$ from opposite ends) to enhance phase conjugation. The slab is assumed to have a cubic nonlinear response with dielectric function $\epsilon_{non}$. The adjacent media on the left and right are
	 linear with dielectric constant $\epsilon_i$. The signal and phase conjugated wave amplitudes are denoted by $A_3$ and $A_4$, respectively.} 
\label{f0}
\end{figure}

We adhere to the standard degenerate four-wave mixing (DFWM) for phase conjugation, which is shown in Fig.~\ref{f0}. Let the nonlinear slab have a cubic response with dielectric function $\epsilon_{non}=\bar{\epsilon}+\alpha|E|^2$ , where $\bar{\epsilon}$ gives the linear background and $\alpha$ the nonlinearity constant.
We assume the pump waves to be much stronger that the the signal and the phase conjugated waves, so that the undepleted pump approximation can be used. The theory for such DFWM has been worked out in detail \cite{yariv-pc,boyd}, and the experimental results are also available \cite{ep-pc}. A very interesting extension was by Kaplan \textit{et al.}, where, the vectorial character of the waves were incorporated, and various eigen polarizations were obtained  \cite{kaplan}. The same study reports the occurrence of the isolas (isolated loops in the input-output characteristics). In order to explore the CPA induced effects we ignore the vector character, and deal with a scalar theory as in Ref.~\cite{yariv-pc}. We also report bistable response that indicate the emergence of the isolas.
\par%
In the undepleted pump approximation the spatial evolution of the pump waves is completely delinked from those of the signal and phase conjugated waves. The evolution of the pump can be easily captured when both the pump waves have the same intensity. 
In that case the symmetry of the structure as regards the pump can be exploited in order to arrive at the dispersion relations for symmetric and anti-symmetric CPA profiles given by \cite{knr2}
\begin{eqnarray}
D_S=p_{1z}+ip_{2z}\tan{\left(p_{2z}\bar{d}/2\right)}=0,\label{eq1}\\
D_A=p_{1z}-ip_{2z}\cot{\left(p_{2z}\bar{d}/2\right)}=0,\label{eq2}
\end{eqnarray}
where the subscript $S$ and $A$ refer to the symetric and anti-symmetric states. Note that the dispersion relations for CPA are different from those of the standard waveguide modes. Recall that for obtaining the former one demands null scattering for non-zero input waves and, for the latter, one solves for maximum scattering for null input \cite{knr2,sdg-wo}. Other parameters in Eqs.~(\ref{eq1})-(\ref{eq2}) are as follows
\begin{eqnarray}
p_{1z}&=&\sqrt{\epsilon_i},\label{eq3}\\
p_{2z}&=&\sqrt{\bar{\epsilon}+3U_2},\label{eq4}
\end{eqnarray}
where $U_2=\alpha |A_2|^2$ is the dimensionless intensity of the waves inside the nonlinear slab and $\bar{d}$ ($\bar{d}=k_0d,~k_0=2\pi/\lambda$) is the dimensionless width of the slab. As can be seen from Eq.~(\ref{eq4}), the normalized propagation constant in the nonlinear slab depends on the incident pump power \cite{felber,sdg-ncm-rev}. Thus the location and the distortion of the CPA dips can be controlled completely by the incident power. Under illumination by pump waves with identical powers, the magnitude of the forward and the backward wave amplitudes inside the slab are the same \cite{knr2}. The same can be used as a parameter to obtain the nonlinear response of the CPA system. An increase in this parameter initially leads to the super-scattering (SS) when most of the energy is scattered out (very large value of $A_f$), to the CPA state ($A_f \approx 0$), when most of incident energy is absorbed.
\par
Having understood the pump induced changes inside the nonlinear slab, one can study the spatial evolution of the signal and the phase conjugated waves governed by \cite{yariv-pc} 
\begin{eqnarray}
\frac{dA_3}{dz}&=&i\kappa_1 A_3+  i\kappa_2 A^*_4 \label{eq5} ,\\
\frac{dA_4}{dz}&=&-i\kappa_1 A_4-i\kappa_2 A^*_3 \label{eq6},
\end{eqnarray}
where $|z|\leq d/2$ and $\kappa_1$ and $\kappa_2$ are given by
\begin{eqnarray}
\kappa_1 &=& \frac{3\omega}{ \text{Re}(\sqrt{\bar{\epsilon}}) c}  (2\alpha|A_2|^2), \label{eq7}\\
\kappa_2 &=& \frac{3\omega}{ \text{Re}(\sqrt{\bar{\epsilon}}) c}  \alpha  A_2^2.\label{eq8}
\end{eqnarray}
In writing the above equations, we have assumed that for symmetric incidence we satisfy the dynamical phase matching condition \cite{boyd}. As shown in Refs.~\cite{yariv-pc,boyd}, Eqs.~(\ref{eq5})-(\ref{eq6}) represent a linear system and can be easily solved for the spatial evolution of the amplitudes $A_3$ and $A_4$ in the nonlinear slab. For typical boundary conditions, say $\sqrt{\alpha}A_3(-d/2)=0.01$ and $\sqrt{\alpha}A_4(d/2)=0$ (these conditions imply that the length $d$ is sufficiently large so that the phase conjugated wave is assumed to have null amplitude at the right interface) one has the phase conjugated amplitude (at the left interface $z=-d/2$) to be \cite{yariv-pc,boyd}
\begin{eqnarray} \label{eq9}
A_4(z=-d/2) &=&  \frac{i\kappa_2}{|\kappa_2|}  \tan \left(|\kappa_2| d \right) A_3^*(-d/2),
\end{eqnarray}
and the phase conjugation reflectivity $R_p$ is given by
\begin{equation}
R_p=\left|\frac{A_4(-d/2)}{A_3^*(-d/2)} \right|^2=\left|i \left(\frac{\kappa_2}{|\kappa_2|} \tan{|\kappa_2|d} \right)\right|^2.\label{eq10}
\end{equation}
Note that $\kappa_2$ is now pump power dependent (see Eq.~(\ref{eq8})) and $|\kappa_2| d=(2n+1)\pi$ implies an instability, termed in the literature as oscillation condition \cite{yariv-pc}. Indeed one can manipulate the `oscillation' in the system, which can be further intensified by CPA when both the waves are completely absorbed in nonlinear medium.\par
In order to bring out the distinctive signature of CPA we choose to work in the domain $\pi/2<|\kappa_2| d< 3\pi/2$. Throughout our calculations we have chosen $d=14.00~\mu$m, $\epsilon_i=1$. The lossy medium is chosen to be a metal-dielectric composite whose dielectric response ($\bar{\epsilon}$) is evaluated using Bruggeman's formula \cite{opex-12,knr2}. The dielectric host is assumed to be silica ($\epsilon_h=2.25$) without any dispersion while the metal (silver with $f=0.082$ as the volume fraction of inclusion) data is obtained from the proper interpolation of the experimental work of Johnson and Christie \cite{J&C}. In principle, such an inclusion renders the nonlinear coefficient $\alpha$ to be complex having dispersion/absorption \cite{t-mat,jenekhe}. In order to retain the simplicity of the problem we assume that the metal inclusions render only the linear part of the susceptibility to be complex leaving $\alpha$ real.
\par
\begin{figure}[h]
\center{\includegraphics[width=8.5cm]{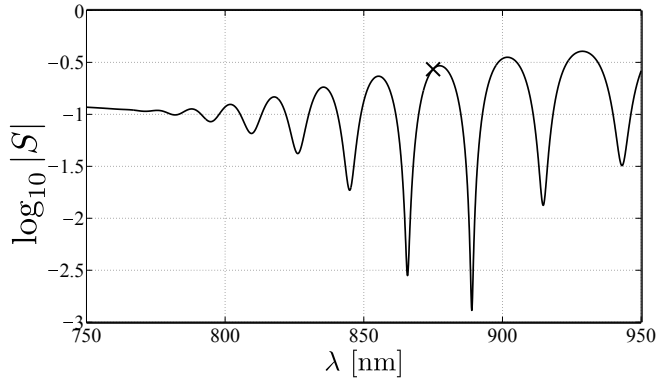}}
\caption{Linear response of the system, $\log_{10}|S|$ as a function $\lambda$. The point marked by the cross ($\lambda=875.0~$nm) denotes the operating point for studying the nonlinear response. The system parameters are: $d=14.00~\mu$m, $\epsilon_h=2.25$, $f=0.082$, $\epsilon_i=1$.} 
\label{f2}
\end{figure}

In Fig.~\ref{f2} we plot the linear response ($\alpha=0$), which shows the typical CPA dip at very low pump powers. It is clear from this figure that one of the Fabry-Perot modes can lead to near perfect absorption. 
\begin{figure}[h!]
	\center{\includegraphics[width=7.0cm]{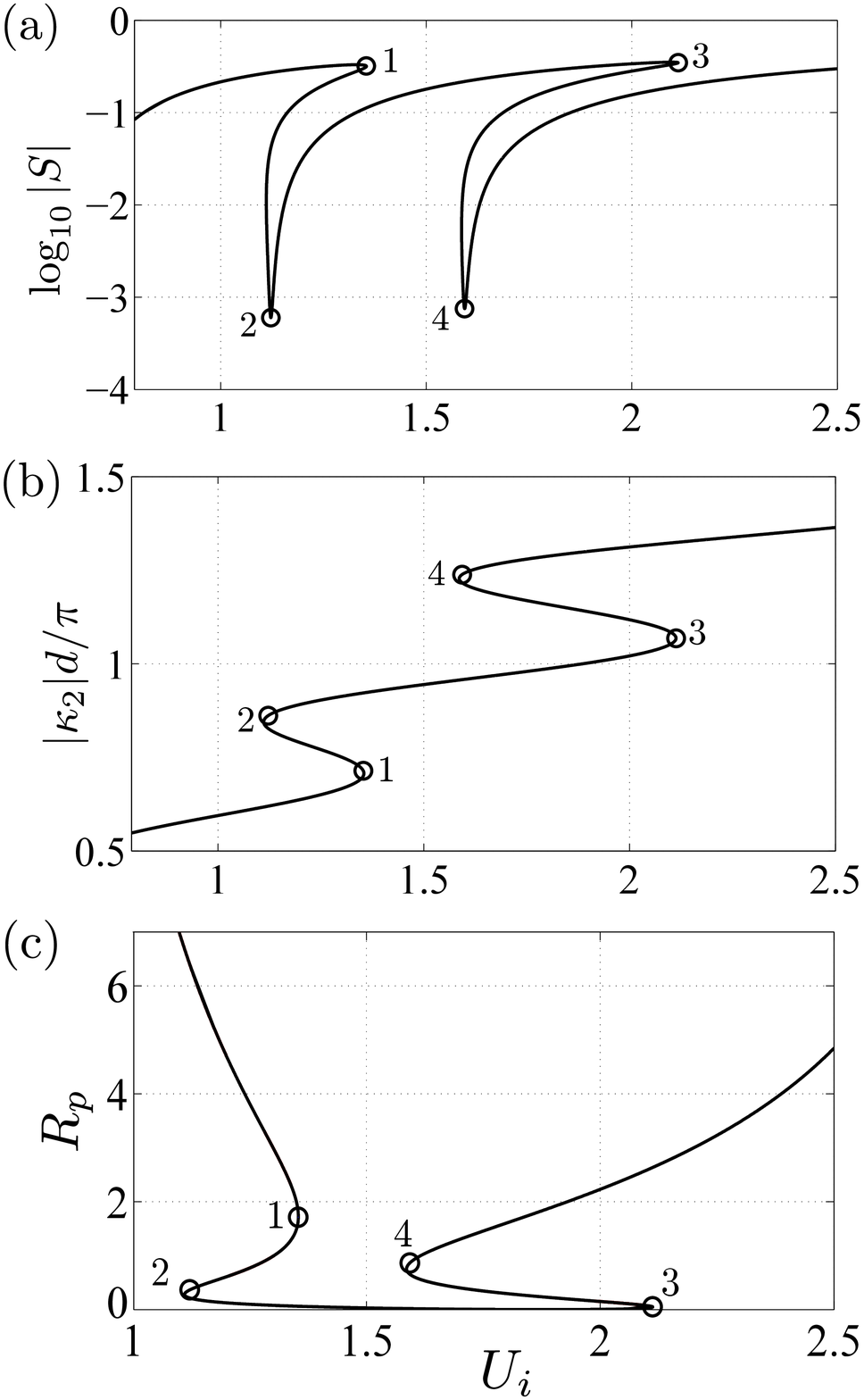}}
	\caption{(a) $\log_{10}|S|$, (b) $|\kappa_2 d|/\pi$ and (c) phase conjugated reflectivity $R_p$ as functions of dimensionless incident intensity $U_i$ for symmetric states. Other parameters are same as in Fig.~\ref{f2}} 
	\label{f3}
\end{figure}
We choose a point on this curve as our operating point at $\lambda=875.0~$nm (point marked by a cross in Fig.~\ref{f2}). The choice is dictated by the fact that an off-resonant system can be brought back to CPA by increasing pump power, while an on-resonant system is taken away from the CPA dip \cite{knr2}. Thus, we choose to work with the detuned linear system in order to tune it back to CPA in the nonlinear regime.
The effect of increasing power on the scattered pump intensity on either side $S=|A_f/A_i|^2$ for the symmetric states (obeying Eq.~(\ref{eq1})) is shown in Fig.~\ref{f3}(a). As reported in literature, one has the bistable response \cite{kaplan, jose}. 
One also has nonlinear CPA, which has been studied earlier in great detail \cite{knr2}. The circular marks in Fig.~\ref{f3} correspond to the bistability thresholds. The upper (lower) threshold correspond to superscattering (CPA) states with maximum in $A_f$ ($A_f\approx 0$). For example, points marked by `1' and `3' (`2' and `4') in Fig.~\ref{f3}(a) correspond to superscattering (CPA) states.  The physics of superscattering and CPA states, makes it clear that for former (latter) one will have lower (higher) energy densities in the nonlinear slab and accordingly lower (higher) values of the coupling constant. The dependence of coupling constant ($\kappa_2$) and the phase conjugation reflectivity ($R_p$) as functions of the dimensionless incident intensity is shown in Fig.~\ref{f3}(b) and Fig.~\ref{f3}(c), respectively. Bistable response repeats in these plots, which is expected. The large phase conjugate reflectivity in  Fig. \ref{f3}(c) at lower input intensities is due to our choice of the operation point close to the oscillation condition $|\kappa_2| d=\pi/2$. The remarkable feature that needs to be noted from Fig.~\ref{f3}(c) is the fact that with increasing input intensities one can have a switch-down (near point `1') or switch-up transition (near point `3') in $R_p$ with both the transitions from the SS states.
\par
\begin{figure}[h]
\center{\includegraphics[width=7.0cm]{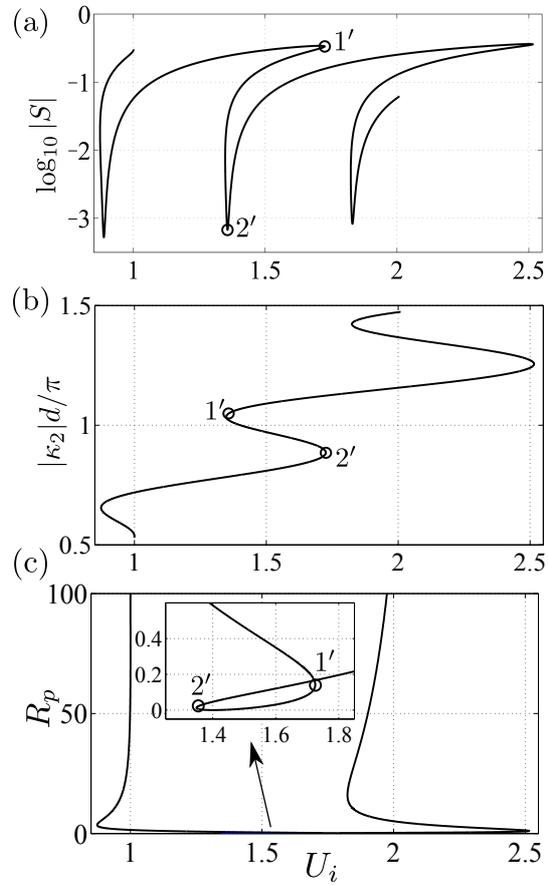}}
\caption{Same as in Fig.~\ref{f3} but now for anti-symmetric states. The inset in Fig.~\ref{f4}(c) depicts the onset of isolas.} 
\label{f4}
\end{figure}

The corresponding results for the anti-symmetric states (obeying Eq.~(\ref{eq2})) are shown in Fig.~\ref{f4}. Results are similar to those in Fig.~\ref{f3} except that one has the precursor to the isolas (see Fig.~\ref{f4}(c)), reported earlier by Kaplan \textit{et al} \cite{kaplan}. 
\par
\begin{figure}[h!]
	\center{\includegraphics[width=8.5cm]{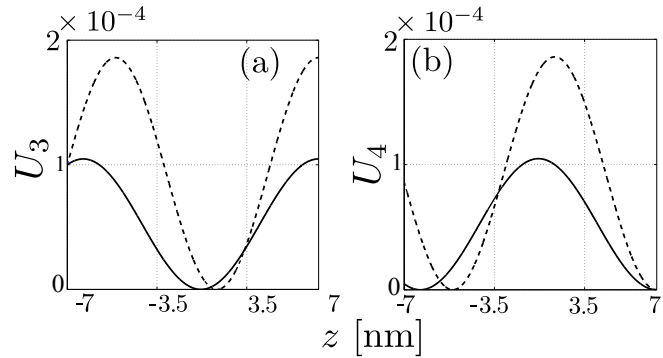}}
	\caption{Dimensionless intensities of (a) signal $U_3$ and (b) phase conjugated $U_4$ waves in the nonlinear slab. The dashed (solid) curves correspond to nonlinear CPA (SS) state marked by point `4' (`3') of Fig.~\ref{f3}.} 
	\label{f5}
\end{figure}
We have also looked at the intensity (dimensionless) distributions of the signal ($U_3=\alpha|A_3|^2$) and phase conjugated ($U_4=\alpha|A_4|^2$) waves inside the nonlinear slab. Fig.~\ref{f5}(a) and Fig.~\ref{f5}(b) depict the variations of $U_3$ and $U_4$, respectively. The dashed and solid curves in Fig.~(\ref{f5}) correspond to SS [labeled `3' in Fig.~(\ref{f3})(c)] and CPA [labeled `4' in Fig.~(\ref{f3})(c)] of the symmetric states, respectively. It can seen that one has a significant variation of $U_3$ and $U_4$ inside the slab when one moves from CPA to SS states. For instance, $U_4$ at the left interface ($z=-7~\mu$m) takes an extremely small value for the SS state, while for the CPA state it is of the order of the incident signal wave intensity. This change in $U_4$ results in a significant increase of the phase conjugate reflectively $R_p$ (see Eq.~(\ref{eq10})). 
\section{Conclusions}
In conclusion, we have investigated phase conjugation in a Kerr nonlinear slab when the counter propagating pump waves are completely absorbed inside the nonlinear medium by coherent perfect absorption. We have made use of the earlier results on light controlled CPA and its ability to switch the system from superscattering to near-total absorption and vice versa \cite{knr2}. We have shown that one can alter the coupling between the signal and the phase conjugated wave with subsequent control on the phase conjugate reflectivity ($R_p$).  Our studies revealed the bi- and multi- stable response in $R_p$ and also the formation of precursor to isolas reported earlier \cite{kaplan}. Moreover we showed that a judicious tuning of the operating point near the oscillation condition can lead to switch-up or switch-down behavior in $R_p$ with increasing input intensities. Our studies can find varied applications for switching, sensing and light-controlled optical systems. 


\end{document}